\documentclass[doublecol,showpacs,amsmath,amssymb]{epl2}
\usepackage{epsfig}
\usepackage{bm}
\title{Full Counting Statistics of Generic Spin Entangler with Quantum Dot-Ferromagnet detectors.}
\author{O. Malkoc, C. Bergenfeldt, P. Samuelsson}

\institute{Department of Physics, Lund University, Box 118, S-221 00
  Lund, Sweden}


\abstract{Entanglement between spatially separated electrons in
  nanoscale transport is a fundamental property, yet to be
  demonstrated experimentally. Here we propose and analyse
  theoretically the transport statistics of a generic spin entangler
  coupled to a hybrid quantum dot-ferromagnet detector system. We show
  that the full distribution of charges arriving at the ferromagnetic
  terminals provide complete information on the spin state of the
  particles emitted by the entangler. This provides means for spin
  entanglement detection via electrical current correlations, with
  optimal measurement strategies depending on the a priori knowledge
  of ferromagnet polarization and spin-flip rates in the detector
  dots. The scheme is exemplified by applying it to Andreev and triple
  dot entanglers.}

\begin{document}

\maketitle 

{\it Introduction -} Entanglement between spatially separated quantum
systems constitutes an indispensable resource for quantum information
processing \cite{Nielsen}. In nanoscale electronic systems, a
promising arena for quantum information and computation, the ultimate
carriers of quantum information are individual electrons. Controlled
creation, spatial separation and detection of entangled electrons thus
constitute key elements in nanoscale quantum information
processing. During the last one and a half decade, a large number of
schemes for transport generation and detection of electronic
entanglement have been proposed \cite{rev1,rev2}. However, a
clear-cut experimental demonstration is still lacking. The main reason
is the paramount difficulty to, in a single nanosystem, generate,
coherently control and unambigously detect the entanglement.

An early key proposal for spin entanglement generation is the quantum
dot based Andreev entangler \cite{AE}; a superconductor coupled to two
quantum dots, further coupled to normal leads. In the transport state
Cooper pairs, electron spin singlets, tunnel out from the
superconductor, via the dots, into the leads. At ideal operation, each
Cooper pair is coherently split without altering the spin properties.
This gives a source of pairs of spatially separated, maximally spin
entangled electrons in the leads. Subsequent theoretical works
extended on or further analysed different properties of the entangler
\cite{AE,Sament,French,French2,Lehur}.

\begin{figure}[h]
\centering
\includegraphics[width=0.45\textwidth]{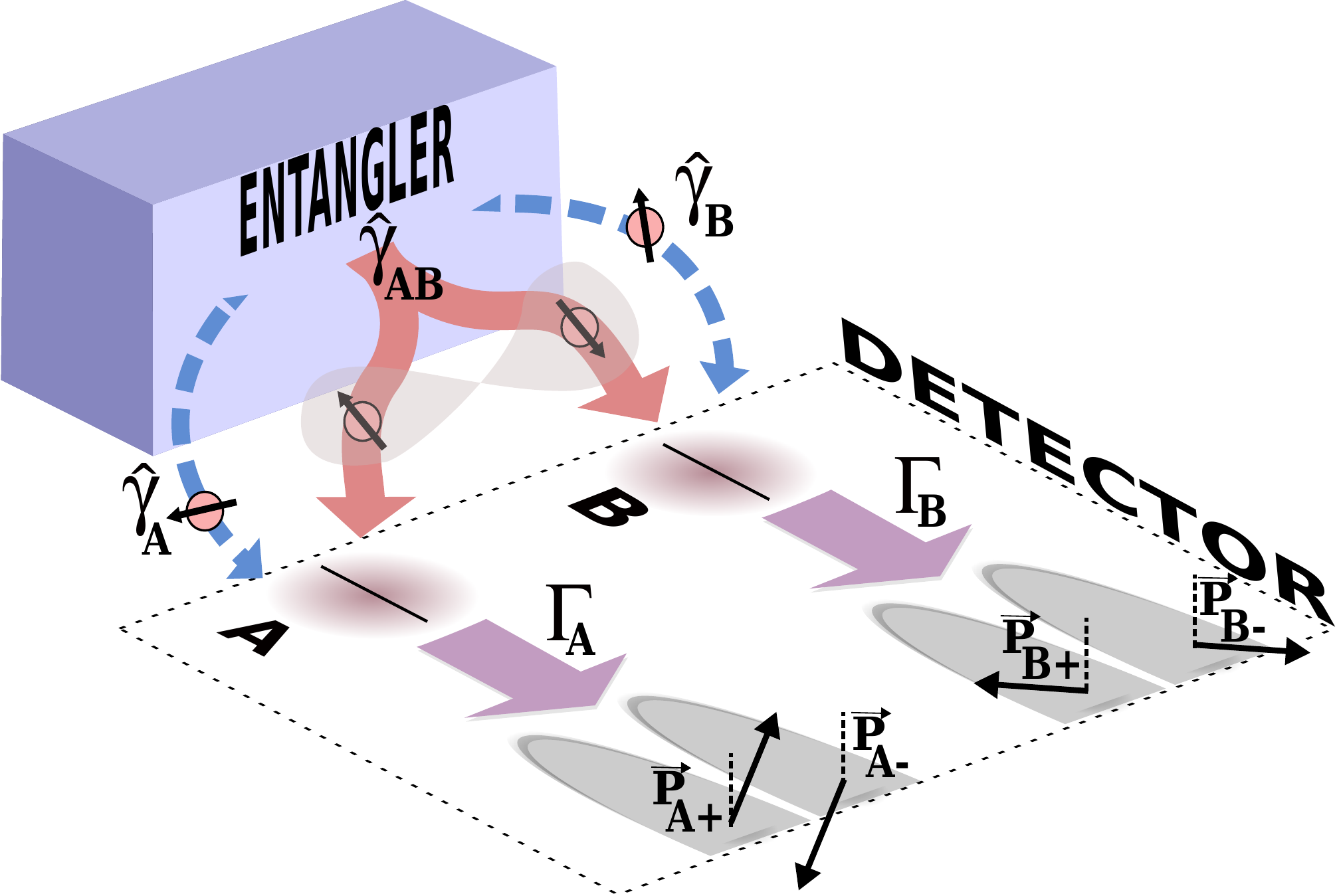}
\caption{Schematic of the combined entangler-detector system. The
  generic entangler emits single (dashed line) or split pairs of
  (solid line) particles, with spin dependent rates described by
  matrices $\hat \gamma_A, \hat \gamma_B$ and $\hat \gamma_{AB}$
  respectively, into the detector quantum dots A and B. The dot A (B), with a
  single, spin degenerate level at energy $\varepsilon_A (
  \varepsilon_B)$, is further tunnel coupled with the same rate
  $\Gamma_A$ ($\Gamma_B$) to two ferromagnetic leads with
  polarisations $\vec{p}_{A+} = -\vec{p}_{A-} ~(\vec{p}_{B+} =
  -\vec{p}_{B-} )$.}
\label{fig:schematic}
\end{figure}

Recently, important steps were taken towards an experimental
demonstration of an Andreev entangler. In a series of experiments
\cite{Schon1,Kontos1,Schon2,Heil,Kontos2,Schon3}, splitting of Cooper
pairs into two quantum dots, formed in semiconductor nanowires or
carbon nanotubes, was reported. Efficient splitting of the pairs was
clearly demonstrated by current\cite{Schon3} and cross correlation
measurements \cite{Heil}. The experiments spurred further theoretical
investigations on various aspects of Cooper pair splitters
\cite{Soller,Rech,Veldhorst,Burset,Hiltscher,Cottet,Braun}.

Importantly, in none of the experiments reported
\cite{Schon1,Kontos1,Schon2,Heil,Kontos2,Schon3} were the spin
properties of the emitted pairs directly investigated. To verify that
the Cooper pair splitters also work as Andreev entanglers, emitting
spin-singlets, non-local spin sensitive detection is necessary. To
this aim, albeit challenging, a natural extension of the experiments
would be to couple the dots to ferromagnetic (FM) leads, see
fig. \ref{fig:schematic}. By performing a set of current cross
correlation measurements \cite{Cht,Sam1,Been,Sam2} with non-collinear
FM-polarization\cite{Cht,Naz} the entanglement can be tested by a Bell
inequality or even quantified by spin state tomography
\cite{tomo,tomoarne}. To facilitate such an experiment under realistic
conditions, in the presence of spurious tunneling processes, spin-flip
scattering in the dots and limited magnitude of the polarization,
several questions need to be carefully adressed. Most importantly i)
how are the spin properties of the pair emitted by the entangler
manifested in the cross correlators of the currents at the FM-leads
and, if possible, ii) how can system parameters and detector settings
be optimized to allow for an unambiguous detection of the entanglement
of the emitted state?

In this work we provide answers to these questions by considering the
full statistics of charge transfer between the entangler and the
FM-leads. To make the scheme applicable beyond Cooper pair splitters
we consider a generic entangler-detector setup, shown in
fig. \ref{fig:schematic}. The entangler emits arbitrary single and
two-particle spin-states into the dots. The two dots together with the
FM-leads constitute the detector. To avoid cross-talk between the two
detector dots as well as back-action of the detector on the entangler
we consider a weak entangler-detector coupling. Importantly, working
within a spin dependent quantum master equation formalism
\cite{Flindt,Kiess,Fazio,Konig} we can treat both charging effects and
spin-flip scattering in the dots, extending on earlier works
\cite{Naz,Morten,Soller} on statistics of entanglers coupled, via a
single non-interacting dot, to FM-electrodes.

The charge transfer statistics allows us to identify the individual
particle tunneling events \cite{Belrew} as well as their spin
properties. Based on this statistics we show how the current cross
correlations provide direct information on the spin properties of the
emitted, entangled two-particle state. In line with earlier work
\cite{Cht,Sam1,Been,Sam2} we find that spurious single particle
tunneling does not affect the correlations. Moreover, depending on how
well the FM-polarizations and the spin-flip rate are characterized, we
propose measurement strategies to optimize the entanglement
detection. To demonstrate the versatility of our scheme we apply it
both to the Andreev entangler\cite{AE} and a triple quantum dot
entangler \cite{Saraga}.

{\it Entangler-Detector system} - The combined entangler-detector
system is shown in fig. \ref{fig:schematic}. The detector subsystem
consists of two quantum dots, A and B, with each dot $\alpha=A,B$
coupled to two FM-leads $\alpha +,\alpha -$ via tunnel barriers with
rates $\Gamma_{\alpha+} = \Gamma_{\alpha-} = \Gamma_{\alpha}/2$. Each
dot has a single spin degenerate level, at energy $\varepsilon_A$ and
$\varepsilon_B$ respectively. Double occupancy of dot A or B is
prevented by strong on-site Coulomb interaction. The FM-leads have
polarisations $\vec p_{\alpha +} = -\vec p_{\alpha -} = \equiv p \vec
n_{\alpha}$ with identical magnitude $p$ and unit vectors $\vec n_A$
and $\vec n_B$ non-collinear.

The generic entangler is acting as a source of both single electrons
and split pairs of spin-correlated electrons, see
fig. \ref{fig:schematic}. The pair emission process is characterized
by a $4\times 4$ rate matrix $\hat \gamma_{AB}$, with elements
$\gamma^{\sigma \sigma', \tau \tau'}_{AB}$ where $\sigma,\sigma',\tau,
\tau' = \uparrow, \downarrow$. This describes emission of pairs with a
spin density matrix $\hat \gamma_{AB}/\mbox{tr}[\hat \gamma_{AB}]$ at
a rate $\mbox{tr}[\hat \gamma_{AB}]$. As a key example, emission of
spin singlets
$|\Psi_S\rangle=(|\!\!\uparrow_A\downarrow_B\rangle-|\!\!\downarrow_A\uparrow_B\rangle)/\sqrt{2}$
with a rate $\gamma$ gives $\hat \gamma_{AB}=\gamma |\Psi_S\rangle
\langle\Psi_S|$. The emission of single particles is correspondingly
described by $2\times 2$ rate matrices $\hat \gamma_{\alpha}$ with
matrix elements $\gamma^{\sigma \sigma'}_{\alpha}$. Throughout the
paper we consider $\gamma_{AB}^{\sigma \sigma', \tau
  \tau'},\gamma_{A}^{\sigma \sigma'},\gamma_{B}^{\sigma \sigma'} \ll
\Gamma_A, \Gamma_B$ so that back-tunnelling from the dots to the
entangler can be neglected. Moreover, we assume $\gamma^{\sigma
  \sigma'}_{\alpha} \leq \gamma_{AB}^{\sigma \sigma', \tau \tau'}$,
achievable for relevant entanglers \cite{AE,Saraga}. In addition, we
account for spin flip scattering in the dots with a rate $\eta$, taken
to be the same for A and B.

The FM-leads are all kept at the same potential. Moreover, a large
bias is applied between the entangler and the FM-leads, in order to
have all detector-entangler energy levels well inside the bias
window. The temperature of the leads is much smaller than the bias as
well as the distance from the detector-entangler energy levels to the
edges of the bias window. This allows us to neglect
back-tunneling from the FM-leads into the dots, known to complicate
the entanglement detection \cite{Titov,Neder}.

{\it Full transport statistics -} As we describe in detail below, in
this high-bias regime the transport properties can be described
exactly within a quantum master equation approach to the reduced spin
density matrix of the dots. The full distribution of charge
transferred to the FM-leads (during a long measurement time) is
conveniently characterised \cite{Belrew} by a cumulant generating
function $F_{\chi}$ where $\chi =
\{\chi_{A+},\chi_{A-},\chi_{B+},\chi_{B-} \}$ denotes the set of lead
counting fields. To leading order in the rate matrices we find
\begin{eqnarray}
\label{fcs}
&& F_\chi =  \sum_{\alpha, m } \mbox{Tr}\left[ \hat Q_{\alpha m}^{\zeta} \hat \gamma_\alpha \right]\left( e^{i\chi_{\alpha m}} -1 \right) \\
 & + & \sum_{n,m} \mbox{Tr} \left[ (\hat Q_{A  n}^{\zeta} \otimes \hat Q_{B m}^{\zeta})   \hat  \gamma_{AB}  \right] \left( e^{i(\chi_{A n} + \chi_{B m})} - 1\right). \nonumber
\end{eqnarray}
where $\otimes$ denotes the direct product, $\hat Q_{\alpha
  m}^{\zeta}=(1/2)[\hat 1+\zeta_{\alpha}\vec n_{\alpha m}\cdot \vec
\sigma]$ is a $2\times 2$ detector matrix with $\vec \sigma=(\hat
\sigma_x,\hat \sigma_y,\hat \sigma_z)$ a vector of Pauli matrices and
$0 \leq \zeta_{\alpha} \leq 1$ the detection efficiency. The
efficiency $\zeta_{\alpha}=p_{\alpha}(1-\eta_{\alpha})$ is a product
of the FM-lead polarization $p_{\alpha}$ and $1-\eta_{\alpha}$, where
$\eta_{\alpha}=\eta/(\Gamma_{\alpha}+\eta)$ is the dimensionless
spin-flip rate in dot $\alpha$, ranging from $0$ for negligible
spin-flip scattering to $1$ for complete spin randomization.

Eq. (\ref{fcs}) is the key technical result of our paper. It allows
for a compelling and physically clear picture of the transport
statistics through the entangler-detector system and provides means to
identify the spin properties of the emitted pairs. The generating
function $F_{\chi}$ in Eq. (\ref{fcs}) describes a set of independent
Poisson
transfer processes of single and pairs of particles: \\
$\bullet$ Each term $\propto 1-e^{i(\chi_{A m}+\chi_{Bn})}$ describes
a pair of particles arriving, one particle to lead $Am$ and one to
$Bn$, with a transfer rate $\mbox{Tr}[(\hat Q_{A m}^{\zeta} \otimes
\hat Q_{B n}^{\zeta})\hat \gamma_{AB}]$. The transfer rate depends on
the spin properties of the emitted pair, via the rate matrix $\hat
\gamma_{AB} $. In particular, for an emitted singlet $\hat
\gamma_{AB}=\gamma|\Psi_S\rangle \langle \Psi_S|$ we have the
two-particle rate $(\gamma/4)[1-\zeta_A\zeta_B\vec n_{Am}\cdot \vec
n_{Bn}]$. This rate is dependent on the relative orientation of the
polarizations via $\vec n_{Am}\cdot \vec n_{Bn}$, clearly
demonstrating the non-local
character of the spin-correlations. \\
$\bullet$ Each term $\propto 1-e^{i\chi_{\alpha m}}$ describes a
single particle arriving at terminal $\alpha m$, with a transfer rate
$\mbox{Tr}[\hat Q_{\alpha m}^{\zeta}\hat \gamma_\alpha]$. Similar to
the two-particle term, the transfer rate depends on the spin properties
of the emitted particle via $\hat \gamma_{\alpha}$.

{\it Detector efficiency vs entanglement suppression -} As is clear
from Eq. (\ref{fcs}), both finite spin-flip scattering $\eta_{\alpha}
>0$ and non-unity polarization $p_{\alpha}<1$ lead to a reduced
detector efficiency $\zeta_{\alpha}<1$. Importantly, the transfer
rates in $F_{\chi}$ can be rewritten as follows, providing a different
picture: Making use of the formal quantum operation approach
\cite{Nielsen} we can write the detector matrix as $\hat Q_{\alpha
  m}^{\zeta}=\mathcal{E}_{\alpha}(\hat Q_{\alpha m})$ where
$\mathcal{E}_{\alpha}(\hat q)=\zeta_{\alpha} \hat
q+(1-\zeta_{\alpha})\mbox{Tr}[\hat q]\hat 1/2$ is the depolarization operation for a
$2\times 2$ matrix $\hat q$ and $\hat Q_{\alpha m}=(1/2)[\hat 1+\vec
n_{\alpha m}\cdot \vec \sigma]$ the ideal detector matrix, for
efficiency $\zeta_{\alpha}=1$. Noting that we can write
$\mathcal{E}_{\alpha}(\hat q)=[1/4](1+3\zeta_{\alpha})\hat
q+[1/4](1-\zeta_{\alpha})(\hat \sigma_x \hat q \hat \sigma_x+\hat
\sigma_y \hat q \hat \sigma_y+\hat \sigma_z \hat q \hat \sigma_z)$, we
can write the single particle transfer rate as
$\mbox{Tr}[\mathcal{E}_{\alpha}(\hat Q_{\alpha m})\hat
\gamma_\alpha]=\mbox{Tr}[\hat Q_{\alpha m}\mathcal{E}_{\alpha}(\hat
\gamma_\alpha)]$, describing perfect detection of a depolarized rate
matrix $\hat \gamma_\alpha$. This is readily extended to the
two-particle transfer rate, which can be written
\begin{equation}
 \mbox{Tr}[\mathcal{E}(\hat Q_{A m} \otimes
  \hat Q_{B n})\hat \gamma_{AB}]= \mbox{Tr}[(\hat Q_{A m} \otimes
  \hat Q_{B n})\mathcal{E}(\hat \gamma_{AB})]
\end{equation}
where $\mathcal{E}=\mathcal{E}_A\otimes\mathcal{E}_B$ describes two
independent, local depolarization operations. For clarity, the
depolarized two-particle rate matrix can be written explicitly
\begin{eqnarray}
  &\!\!\!\!\!\!\!\!\!&\mathcal{E}(\hat\gamma_{AB})=\zeta_A\zeta_B\hat \gamma_{AB}+ \frac{\zeta_A(1-\zeta_B)}{2} \mbox{Tr}_B[\hat\gamma_{AB}]\otimes \hat 1 \nonumber \\
  &\!\!\!\!\!\!\!\!\!\!+&\!\!\!\!\!\! \frac{\zeta_B(1-\zeta_A)}{2}  \hat 1\otimes \mbox{Tr}_A[\hat\gamma_{AB}]+\frac{(1-\zeta_A)(1-\zeta_B)}{4} \hat 1\otimes \hat 1 
\end{eqnarray}
where $\mbox{Tr}_{\alpha}[..]$ denotes a partial trace over the spin
degrees of freedom in dot $\alpha$. Taking again the example of
(maximally entangled) spin singlets emitted with a rate $\gamma$,
i.e. $\hat \gamma_{AB}=\gamma |\Psi_S\rangle \langle \Psi_S|$, the
depolarized rate $\mathcal{E}(\hat \gamma_{AB})=\gamma
[\zeta_A\zeta_B|\Psi_S\rangle \langle \Psi_S|+ [1/4](1-\zeta_A\zeta_B)
\hat 1\otimes \hat 1]$ describes emission of Werner states
\cite{Werner}, entangled only for $\zeta_A\zeta_B>2/3$. This clearly
illustrates the following: the two particle transfer rate is the same
for maximally entangled states detected with reduced efficiency as
for partially entangled states detected with unit efficiency. As we
now discuss, this insight greatly helps to develop measurement
strategies for an unambiguous entanglement detection.

{\it Cross correlations and entanglement detection -} From $F_{\chi}$
the different low frequency cumulants are obtained by successive
derivatives with respect to the counting fields. For the average
electrical current at terminal $A m$ (and similarly for $B_n$) we have
$I_{A m } = -i e \partial_{ \chi_{A m}} F_\chi |_{\chi = 0 }$ giving
\begin{equation}
  I_{A m } = e\mbox{Tr}\left [\hat Q_{A m}^{\zeta} (\hat \gamma_A + \mbox{Tr}_{B} \left[\hat \gamma_{AB}\right]) \right].
  \label{currAm}
\end{equation}
The average current provides information about the single particle
processes through A via $\hat \gamma_A$ as well as the local, reduced
single particle properties of the emitted pairs, via $\mbox{Tr}_{B}
\left[\hat \gamma_{AB}\right]$. Consequently, $I_{Am}$ and $I_{Bn}$
are local quantities and can not provide full information on the
emitted two-particle state, in particular not on the entanglement.

Turning instead to the non-local cross correlations between currents
at reservoirs $\alpha m$ and $\beta n$, obtained as $S_{\alpha m ,
  \beta n} = - e^2 \partial_{\chi_{\alpha m}}\partial_{\chi_{\beta n}}
F_\chi|_{\chi = 0} $, we have
\begin{equation}
\label{eq:S1}
  S_{Am, Bn } = e^2\mbox{Tr}\left [(\hat Q_{A m}^{\zeta} \otimes \hat Q_{B n}^{\zeta}) \hat \gamma_{AB} \right]
\end{equation}
From Eqs. (\ref{eq:S1}) and (\ref{fcs}) it is clear that $S_{Am, Bn }$
is directly proportional to the corresponding two-particle emission
rate. In particular, $S_{Am, Bn }$ provides direct information about
the spin properties of the individual pairs, via $\hat
\gamma_{AB}$. Moreover, $S_{Am, Bn }$ does not contain any information
about the spurious single-particle emission or correlation between
emitted pairs (contributes only to next order in $\hat
\gamma_\alpha/\Gamma_{\beta}, \hat \gamma_{AB}/\Gamma_{\beta}$). This
illustrates in a compelling way that a long time measurement, with a
large number of emitted pairs collected in the leads, effectively
\cite{Cht,Sam1,Been,Sam2} constitutes an average over a large number
of identically prepared pair spin states. 

Importantly, the form of the cross correlator in Eq. (\ref{eq:S1})
allows in principle for entanglement detection via e.g. \cite{tomo} a
complete tomographic reconstruction of $\hat \gamma_{AB}$ or a test of
a Bell inequality \cite{Cht,Sam1,Been,Sam2}. In both cases, one needs
to perform a set of measurements with different polarization settings
$\vec n_A$ and $\vec n_B$. However, the interpretation of the
measurement result, in particular the answer to the question ``is the
emitted state entangled?'', depends both on the method of detection as
well as an accurate knowledge of the detector efficiencies. This is
clearly illustrated by considering separately
two cases:\\
$\bullet$ When the detector efficiencies $\zeta_{\alpha}$ are
accurately known, i.e. both the FM-polarizations $p_{\alpha}$ and the
spin-flip rates $\eta_{\alpha}$ can be faithfully determined, a
quantum spin tomography is in principle viable \cite{tomoarne} for
arbitrary $\zeta_{\alpha}$. In contrast, a Bell inequality test can
only be performed for a limited range of efficiencies. Interestingly,
as was discussed by Eberhardt already two decades ago \cite{Eber}, an
a priori knowledge about $\zeta_{\alpha}$ allows one to optimize
polarization settings $\vec n_{\alpha}$, increasing the efficiency
range for which a Bell inequality violation
is possible. \\
$\bullet$ When the efficiencies $\zeta_{\alpha}$ are not known, a
quantum spin tomography can give an incorrect two-particle state. In
particular, the reconstructed state can have an entanglement larger
than the emitted state, opening up for an incorrect conclusion that
entanglement has been detected. This ``false detection'' scenario can
be illustrated by considering emission of Werner states $\kappa
|\Psi_S\rangle \langle \Psi_S|+ [1/4](1-\kappa) \hat 1\otimes \hat
1$. An underestimation of the detector joint efficiency
$\zeta_A\zeta_B$ by a factor $1/\kappa$ will then lead to
tomographically reconstructed singlet state $|\Psi_S\rangle \langle
\Psi_S|$, maximally entangled. In contrast, a Bell test with unknown
detector efficiencies can not lead to ``false detection'' of
entanglement \cite{comm}. However, for unknown or ill-characterized
efficiencies it is difficult to identify detector settings for an
optimal violation, making a Bell test experimentally more demanding.

{\it Quantum master equation -} We now turn to the derivation of the
transport statistics, in terms of the reduced density operator of the
state in the dots, $\rho = \rho(t)$. In the high-bias limit under
consideration, the dynamics of $\rho$ can be described exactly by a
Liouville equation on Lindblad form
\begin{equation}
\label{eq:eom}
\frac{d \rho}{dt}  = \mathcal{L}_{H}(\rho)+ \mathcal{L}_{1}( \rho ) + \mathcal{L}_{2}( \rho )+\mathcal{L}_{\eta}( \rho ) + \mathcal{L}^\chi_{FM}(\rho).
\end{equation}
Here the term $\mathcal{L}_{H}(\rho)=- \frac{i}{\hbar}[H_{d}, \rho]$
describes the free evolution of the dot state, with $H_{d} =
\sum_{\alpha \sigma} \varepsilon_{\alpha} d_{\alpha \sigma}^\dagger
d_{\alpha \sigma} $ and $ d_{\alpha \sigma}^\dagger ( d_{\alpha
  \sigma})$ creating (annihilating) electrons in dot $\alpha$, with
spin $\sigma$. The terms $\mathcal{L}_1(\rho)$ and
$\mathcal{L}_2(\rho)$ describe the injection, from the entangler to
the dots, of single and two-particle states respectively and are given
by
\begin{equation} 
\mathcal{L}_{1}(\rho) = \displaystyle\sum \limits_{\alpha \sigma \sigma'} \gamma_\alpha^{\sigma \sigma'}  \left [   d^\dagger_{\alpha \sigma} \rho d_{ \alpha \sigma'} -  \frac{1}{2}  \{   d_{\alpha \sigma'} d^\dagger_{\alpha \sigma}, \rho\}   \right]
\end{equation}
and, 
\begin{eqnarray}
\mathcal{L}_{2}(\rho) &= & \sum \limits_{\tau \sigma \tau' \sigma'} \gamma_{AB}^{\sigma \sigma', \tau \tau'}   \left [ d^\dagger_{A\sigma} d^\dagger_{B \tau} \rho d_{B \tau'} d_{A \sigma'}  \right.  \\ \nonumber
 &   - & \left.   \frac{1}{2}  \{d_{B \tau'} d_{A \sigma'} d^\dagger_{A \sigma} d^\dagger_{B \tau}, \rho   \} \right ].  
\end{eqnarray}
To preserve the trace and ensure positivity of the density matrix the
emission rate matrices must be Hermitian $\hat \gamma_\alpha = \hat
{\gamma}_\alpha^\dagger$, $\hat \gamma_{AB} =
\hat{\gamma}_{AB}^\dagger$. Entangler examples with detailed
derivations of the one and two-particle rate matrices are given
below. Spin flip scattering in the dots, with a rate $\eta$, is
accounted for by the term
\begin{equation}
\mathcal{L}_{\eta}(\rho)=\eta \sum_{\alpha \sigma} \left[  d_{\alpha\sigma}\rho d_{\alpha\sigma}^{\dagger}-\frac{1}{2}\{d_{\alpha\sigma}^{\dagger}d_{\alpha\sigma},\rho\}\right] 
\end{equation}

The last term $\mathcal{L}_{FM}^{\chi}(\rho)$ accounts for the
coupling to the FM-reservoirs.  In order to describe the full charge
transfer statistics we have included counting
fields $\chi_{\alpha m}$, with $m = \pm$, in the terms describing
tunnelling out to the FM-reservoirs. This gives
\begin{eqnarray}
\mathcal{L}^\chi_{FM}(\rho) &=& \sum \limits_{\alpha m \sigma' \sigma}\Gamma_\alpha \bigg [d_{\alpha \sigma} \rho d^\dagger_{\alpha \sigma'}  Q^{\sigma' \sigma}_{\alpha m } e^{i \chi_{\alpha m}} \nonumber \\
&-& \frac{1}{2} \{ d^{\dagger}_{\alpha \sigma} d_{\alpha  \sigma}, \rho \} \bigg  ]
\end{eqnarray}
where $Q^{\sigma \sigma'}_{\alpha m }=(\hat Q_{\alpha m
})_{\sigma\sigma'}$.

Working in the local spin-Fock basis
$\{|0\rangle,|\sigma_A\rangle,|\tau_B\rangle,|\sigma_A\tau_B\rangle\}$,
with $\sigma,\tau=\uparrow,\downarrow$, Eq. (\ref{eq:eom}) can be
written as a linear matrix equation $\partial_t \vec \rho_\chi = \hat
{\mathcal M}_\chi \vec \rho_\chi$. Here $\hat {\mathcal M}_\chi$ is a
$\chi$-dependent transition rate matrix and the ($\chi$-dependent)
vector $\vec \rho_\chi = [\rho_0, \vec \rho_A, \vec \rho_B, \vec
\rho_{AB}]$, where $\rho_0$ is the matrix element for both dots empty
and $\vec \rho_\alpha (\vec \rho_{AB})$ a vector with the elements for
only dot $\alpha$ (both dot A and B) occupied, including one (two)
particle spin coherences.

Following ref. \cite{Flindt} the generating function $F_{\chi}$ can then be obtained from the eigenvalue problem 
\begin{equation}
\hat {\mathcal M}_\chi \vec \rho_\chi  = F_\chi \vec \rho_\chi,
\label{eigenvalueeq}
\end{equation}
To leading order in $ \gamma^{\sigma \sigma'}_{\alpha}/\Gamma_{\alpha}, \gamma^{\sigma \sigma', \tau \tau'}_{AB}/ \Gamma_{\alpha}$, it is possible to solve Eq. (\ref{eigenvalueeq}) analytically, giving Eq. (\ref{fcs}) above.

{\it Transfer rate matrices -} We now turn to a discussion of the
single and two-particle transfer rate matrices $\hat \gamma_{\alpha}$
and $\hat \gamma_{AB}$. As pointed out above, we consider an entangler
subsystem which is weakly coupled to the dots A and B. Together with
the high bias limit, this implies that single and pairs of particles
which have tunneled out of the entangler will only tunnel out to the
FM-leads, and never back to the entangler. In addition, we make the
assumption that the many-body state of the entangler has a well
defined energy $E_e$. We can then evalute the rate matrices within a
T-matrix formulation of time-dependent many-body perturbation theory
\cite{Karsten}. Discussing explicitly the key quantity, the
two-particle rate matrix $\hat \gamma_{AB}$, we have the spin
dependent golden rule result
\begin{equation}
\hat \gamma_{AB}=\hat T\frac{\Gamma_A+\Gamma_B}{(\varepsilon_A+\varepsilon_B-E_e)^2+(\Gamma_A+\Gamma_B)^2/4}.
\label{twopart}
\end{equation}
The spin dependence is contained in the matrix $\hat T$, which has
elements 
\begin{equation}
(\hat T)_{\sigma\sigma',\tau\tau'}=\mbox{Tr}\{\rho_eH_T^{(2)}|\sigma_A\tau_B\rangle\langle
\tau'_B\sigma'_A|H_T^{(2)}\},
\end{equation}
where $\rho_e$ is the density matrix of the isolated entangler,
$H_T^{(2)}$ the effective two-particle entangler-dot tunneling
Hamiltonian and the trace is running over the degrees of freedom of
both the entangler and the dots. The second factor in
Eq. (\ref{twopart}) is $\int dE_A dE_B
\nu_A(E_A)\nu_B(E_B)\delta(E_A+E_B-E_i)$ where
$\nu_{\alpha}(E_{\alpha})=(\Gamma_{\alpha}/\pi)[(E_{\alpha}-\epsilon_{\alpha})^2+\Gamma_{\alpha}^2/4]^{-1}$
is the density of states of dot $\alpha=A,B$, broadened by the
coupling to the FM-leads. The single particle rate matrices $\hat
\gamma_{\alpha}$ can be calculated in a similar (and simpler)
manner. Note that in evaluating Eq. (\ref{twopart}), the polarization
and tunnel rate conditions $\vec p_{\alpha+}=-\vec p_{\alpha-}$ and
$\Gamma_{\alpha+}=\Gamma_{\alpha-}$ allow us to treat the two
FM-reservoirs coupled to dot $\alpha$ as one effective, normal
reservoir. For the same reason, spin-flip scattering in the dots have
no effect on the result in Eq. (\ref{twopart}).

The expression in Eq. (\ref{twopart}) opens up for a treatment of a
wide range of two-particle spin entanglers coupled to quantum dot
detectors, including entanglers with a possibly mixed spin state or
spin-dependent entangler-dot tunneling. The only information required
to evaluate the transfer rate matrices is the effective one and
two-particle tunneling rates and the energy and spin-properties of the
isolated entangler state. To demonstrate the viability of our approach
we analyse two proposed quantum dot based spin entanglers.

\begin{figure}[h]
\centering
\includegraphics[width=0.45\textwidth]{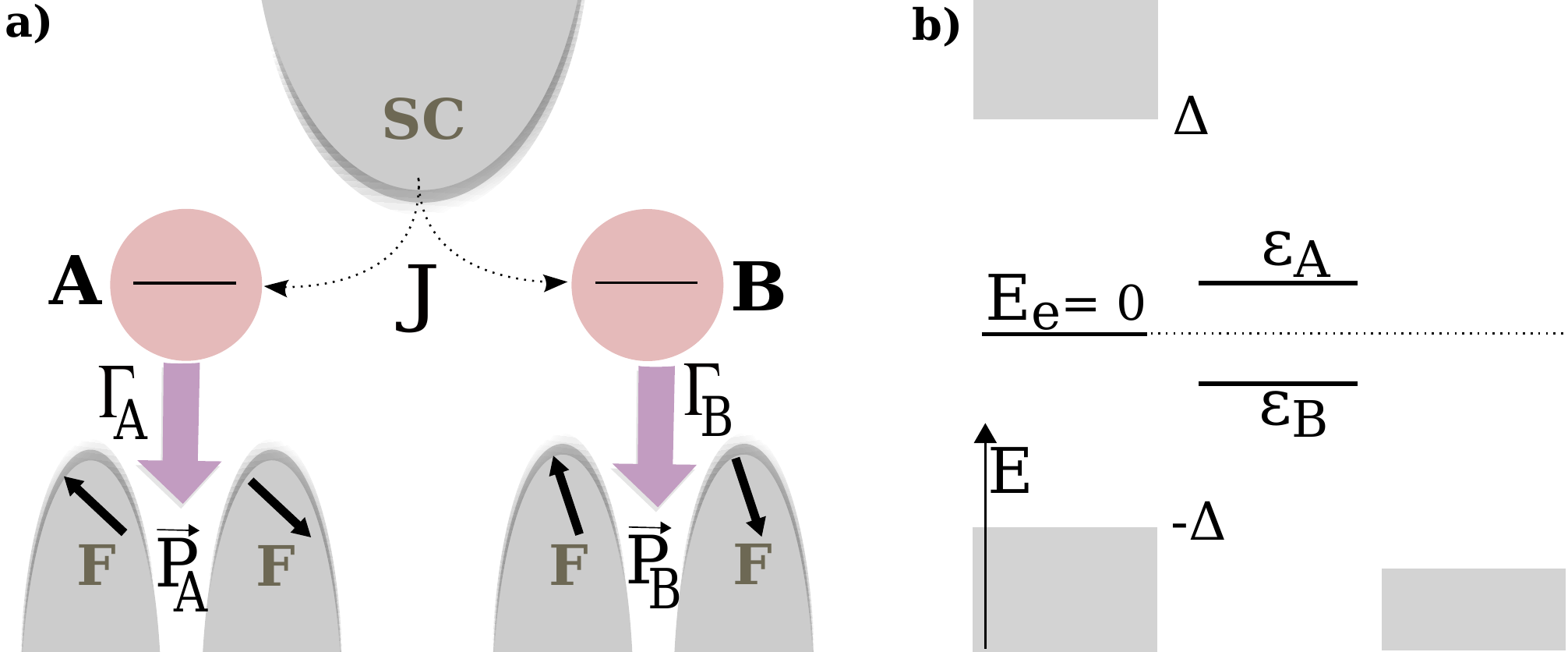}
\caption{a) Schematic of Andreev entangler-detector system. Split
  Cooper pairs are emitted from the superconductor (SC) into dots a
  and B with a rate $J$. For other tunnel rates we refer to
  fig. \ref{fig:schematic}. b) Energy level diagram of the combined
  entangler-detector system, with superconducting gap $\Delta$ and
  ground state $E_e=0$ as well as dot level energies
  $\epsilon_A,\epsilon_B$ shown.}
\label{fig2}
\end{figure}

{\it Andreev entangler -} We first consider an Andreev entangler, of
large interest due to the recent Cooper pair splitter experiments
\cite{Schon1,Kontos1,Schon2,Heil,Kontos2,Schon3}, discussed above. For
completeness we show a schematic of the Andreev entangler-dot detector
system in fig. \ref{fig2}, including relevant energies and tunneling
rates.
In line with our earlier assumptions we here consider the case where
the dominating two-particle process is emission of split Cooper pairs
into the two dots. The processes where the two particles tunnel to the
same dot are suppressed due to large on-site Coulomb
interaction. Moreover, considering dot energies
$\epsilon_A,\epsilon_B$ well inside the superconducting gap $\Delta$
the single particle rates are smaller than or of the order of the
two-particle, pair tunneling rates \cite{AE}, i.e. $\gamma^{\sigma
  \sigma'}_{\alpha} \leq \gamma_{AB}^{\sigma \sigma', \tau \tau'}$.

We recall that the state of the isolated entangler is the
superconducting ground state, with an energy $E_e$ here taken to be
zero. Moreover, the effective two-particle tunneling Hamiltonian can
be conveniently be written \cite{Pistolesi} $H_T^{(2)}=J\left[b_0
  (d_{A \uparrow}^{\dagger}d_{B \downarrow}^{\dagger}-d_{A
    \downarrow}^{\dagger}d_{B \uparrow}^{\dagger})+h.c.\right]$. Here
$J$ is the tunneling element, depending on the properties of the
superconductor and the coupling to the dots, and $b_0$ the destruction
operator of a Cooper pair in the superconductor with the properties
$\langle b_0\rangle=\langle b_0^{\dagger}\rangle=\langle
b_0^{\dagger}b_0\rangle=1$, where the average is taken with respect to
the superconducting ground state. We then directly obtain $\hat
T=J^2|\Psi_S\rangle \langle \Psi_S|$ with $|\Psi_S\rangle$ the spin
singlet state and, writing out explicitly, the two-particle rate
matrix in Eq. (\ref{twopart}) as
\begin{equation}
\hat \gamma_{AB}=\frac{2J^2(\Gamma_A+\Gamma_B)}{(\varepsilon_A+\varepsilon_B)^2+(\Gamma_A+\Gamma_B)^2/4}|\Psi_S\rangle \langle \Psi_S|.
\label{AEgam}
\end{equation}
Along the same lines one can obtain $\hat \gamma_{\alpha}$. With $\hat
\gamma_{AB}$ and $\hat \gamma_{\alpha}$ we can then via
Eq. (\ref{fcs}) evaluate the full, spin dependent transport statistics
for the Andreev entangler. We stress that from the expressions for the
current, Eq. (\ref{currAm}), and cross correlations,
Eq. (\ref{eq:S1}), we reproduce known results \cite{AE,French} in the
parameter limits corresponding to our assumptions.

\begin{figure}[htb]
\centering
\includegraphics[width=0.45\textwidth]{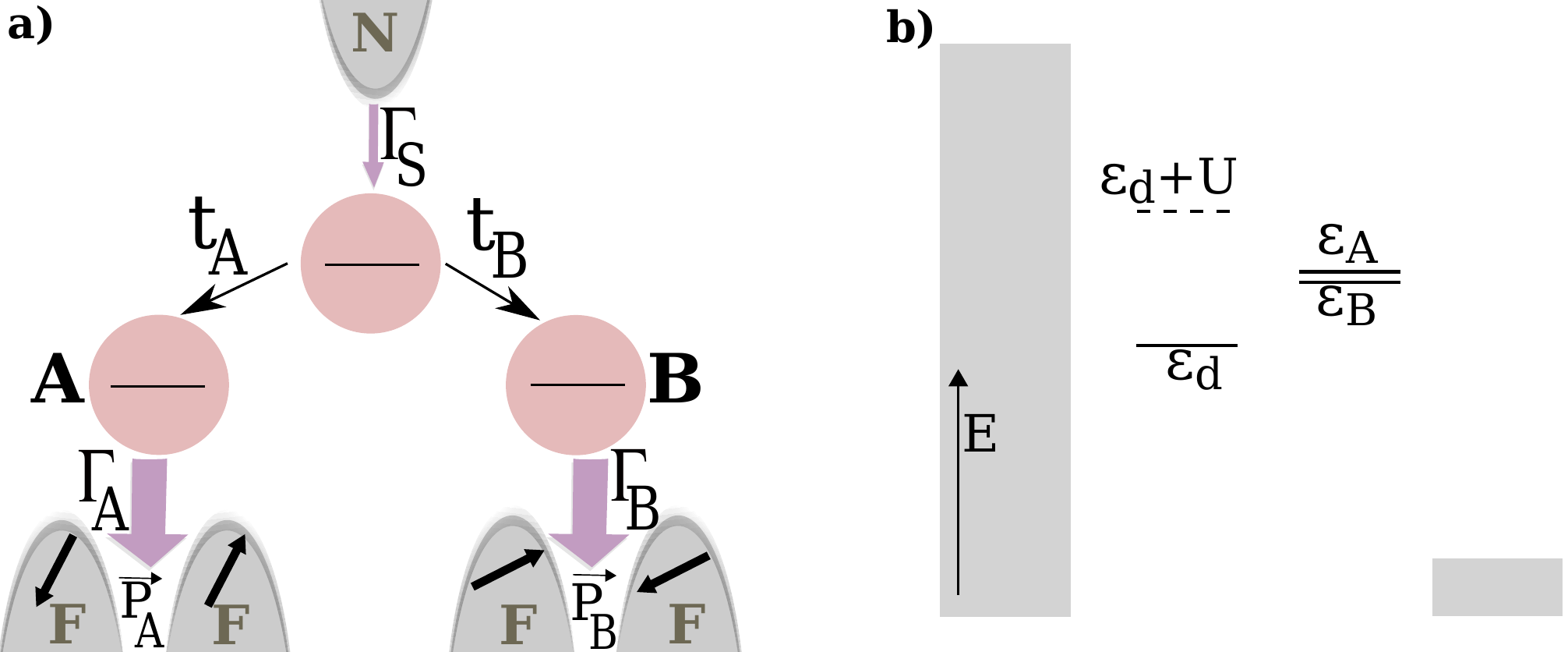}
\caption{a) Schematic of triple dot entangler-detector system. The
  entangler dot is coupled to the detector dots via tunnel barriers
  with rates $t_A,t_B$ and to a normal lead with rate $\Gamma_S$.  For
  other tunnel rates we refer to fig. \ref{fig:schematic}. b) Energy
  level diagram of the combined entangler-detector system, with
  $\epsilon_A,\epsilon_B$ the detector dot level energies,
  $\epsilon_d$ the entangler dot level energy and $U$ the entangler
  dot on-site Coulomb interaction strength. At $2\epsilon_d+U\approx
  \epsilon_A+\epsilon_B$ two particles tunnel resonantly from the
  entangler dot to the detector dots A and B, with an effective rate
  $\propto t^2/U$.}
\label{fig3}
\end{figure}

{\it Triple dot entangler -} As a second example we consider
triple-dot entangler proposed in ref. \cite{Saraga}. Here the
entangler consists of a quantum dot with a single, spin degenerate
level at energy $\epsilon_d$ and an on-site interaction strength $U$,
coupled to a normal lead. The entangler dot is further coupled to the
detector dots via tunnel barriers with rates $t_A=t_B=t$. A schematic
of the entangler-detector system is shown in fig. \ref{fig3}. As
stated above, double occupancy of the detector dots A and B is
prohibited by strong on-site interactions. To have a two particle rate
larger than or of the order of the single particle rates the level
energies $\epsilon_d,\epsilon_A$ and $\epsilon_B$ are tuned to meet
the two-particle resonance condition $2\epsilon_d+U
\approx\epsilon_A+\epsilon_B$. Moreover, single particle resonances,
$\epsilon_{\alpha}\approx \epsilon_d, \epsilon_d+U$, are avoided.

Under these assumptions, together with the high bias condition, the
state of the entangler dot is simply the double occupied level
$d_{e\uparrow}^{\dagger}d_{e\downarrow}^{\dagger}|0\rangle$, with
$d_{e\sigma}^{\dagger}$ creating an electron with spin $\sigma$ in the
entangler dot. The effective two-particle tunneling Hamiltonian is
further given by
$H_T^{(2)}=[2t^2/U](d_{e\uparrow}d_{e\downarrow}[d_{A\uparrow}^{\dagger}d_{B\downarrow}^{\dagger}-d_{A\downarrow}^{\dagger}d_{B\uparrow}^{\dagger}]
+h.c.)$, where in the prefactor $t^2$ comes from the two
entangler-detector dot tunnel events and $U$ is the "energy cost" for
the virtual state created by the first particle, tunneling off
resonance. The two-particle rate matrix in Eq. (\ref{twopart}) then
becomes
\begin{equation}
\hat \gamma_{AB}=\frac{8t^4}{U^2}\frac{\Gamma_A+\Gamma_B}{\delta \epsilon^2+(\Gamma_A+\Gamma_B)^2/4}|\Psi_S\rangle \langle \Psi_S|.
\label{TDEgam}
\end{equation}
where $\delta \epsilon=2\epsilon_d+U-(\epsilon_A+\epsilon_B)$, the energy away from two-particle resonance. 
Along the same lines one can obtain $\hat \gamma_{\alpha}$. From the expressions for the current,
Eq. (\ref{currAm}), we reproduce known results \cite{Saraga}
in the parameter limits corresponding to our assumptions.

{\it Conclusions -} We have investigated the full counting statistics
of spin dependent single and two-particle transfer in a generic
entangler coupled to quantum dot-ferromagnet detectors. From the full
statistics we have identified individual charge transfer
events. Moreover, we have demonstrated that the current cross
correlators can be used to determine the two-particle spin state even
in the presence of spin-flip scattering and limited ferromagnet
polarization. For the future, it would be interesting to investigate
in more detail how the obtained results are modified when relaxing
assumptions such as equal dot-ferromagnet couplings and ferromagnet
polarizations as well as single occupancy of detector dots, in order
to strengthen the connection to experimentally realizable systems. We
note that during the preparation of the present manuscript we became
aware of the recent work \cite{new} where related aspects of
entanglement detection in Cooper pair splitters were discussed.

{\it Acknowledgements -} We acknowledge support from the Swedish VR.


\begin{thebibliography}{02}
\bibitem{comm} Note that in electronic entanglers there is no
  ``detection loophole'', i.e. all particles contribute to measured
  the cross correlators.
\end{thebibliography}
\end{document}